\documentstyle[psfig]{l-aa}

\begin{document}

\thesaurus{ 06(08.02.6; 08.09.2 (TWA-7); 08.12.1; 08.16.5) }

\title{ On the possibility of ground-based direct imaging detection 
of extra-solar planets: The case of TWA-7\thanks{Based on observations 
obtained at the European Southern Observatory, La Silla 
(ESO Proposals 62.I-0418 and 63.N-0178), 
and on observations made with the NASA/ESA Hubble Space 
Telescope, obtained from the data archive at the Space Telescope Science 
Institute. STScI is operated by the Association of Universities for Research 
in Astronomy, Inc., under the NASA contract NAS 5-26555.} }

\author{ R. Neuh\"auser\inst{1} \and W. Brandner\inst{2} \and A. Eckart\inst{1}
\and E. Guenther\inst{3} \and J. Alves\inst{4} \and T. Ott\inst{1} 
\and N. Hu\'elamo\inst{1} \and M. Fern\'andez\inst{1} }

\offprints{R. Neuh\"auser, rne@mpe.mpg.de }

\institute{MPI f\"ur extraterrestrische Physik, Giessenbachstra\ss e 1, D-85740 Garching, Germany
\and University of Hawaii, Institute for Astronomy, 2680 Woodlawn Dr., Honolulu, HI 96822, USA
\and Th\"uringer Landessternwarte Tautenburg, Sternwarte 5, D-07778 Tautenburg, Germany
\and European Southern Observatory, Karl-Schwarzschild-Stra\ss e 2, D-85748 Garching, Germany
}

\date {Received 24 Sep 1999; accepted 23 Dec 1999}

\maketitle

\markboth{Neuh\"auser et al.: Ground-based direct imaging detection of extra-solar planets}{}

\begin{abstract}

We show that {\em ground-based} direct imaging detection of extra-solar
planets is possible with current technology. As an example, we
present evidence for a possible planetary companion to the 
young T Tauri star 1RXSJ104230.3$-$334014 (=TWA-7), discovered 
by ROSAT as a member of the nearby TW Hya association.
In an HST NICMOS F160W image, an object is detected that is more than 
9 mag fainter than TWA-7, located $2.445 \pm 0.035^{\prime \prime}$ south-east 
at a position angle of $142.24 \pm 1.34^{\circ}$. 
One year later using the ESO-NTT with 
the SHARP speckle camera, we obtained H- and K-band detections of 
this faint object at a separation of $2.536 \pm 0.077^{\prime \prime}$
and a position angle of $139.3 \pm 2.1^{\circ}$.
Given the known proper motion of TWA-7,
the pair may form a proper motion pair. 
If the faint object orbits TWA-7,
then its apparent magnitudes of H=$16.42 \pm 0.11$ and 
K=$16.34 \pm 0.15$ mag yield absolute magnitudes consistent with a
$\sim 10^{6.5}$ yr old $\sim 3$~M$_{\rm jup}$ mass object according to the 
non-gray theory by Burrows et al. (1997). At $\sim 55$~pc, the angular separation 
of $\sim 2.5^{\prime \prime}$ corresponds to $\sim 138$~AU.
However, position angles and separations are slightly more consistent with 
a background object than with a companion. 


\end{abstract}

\section{Introduction: Direct imaging of planets}

Several extra-solar planet candidates 
have been detected indirectly by radial velocity variations of stars 
(review by Marcy \& Butler 1998), 
one such candidate is confirmed by a transit event (Charbonneau et al. 2000).
Direct imaging detection of planets like those in our solar system 
but orbiting other stars is difficult due to the limited dynamical range:
Planets are too faint and too close to bright stars.
Planets are thought to form in circumstellar disks around young stars,
which are typically hundreds of AU in size
(e.g. McCaughrean \& O'Dell 1996).
One can try to avoid the problem of dynamical range by searching for planetary 
companions around nearby stars, where the typical disk size
corresponds to several arc sec, sufficient to resolve a faint object
next to a bright star. However, nearby stars usually are too old, so that 
their planets are too faint for direct detection with current technology. 
Young planets are still self-luminous due to on-going accretion and/or 
contraction (Burrows et al. 1997, Brandner et al. 1997, Malkov et al. 1998)
and sufficiently bright for direct detection.
This should best be possible in the infrared bands H and K, 
where the brightness difference between young stars and 
young planets is expected to be the lowest (Burrows et al. 1997).

A few sub-stellar companions to normal stars were detected already
by direct imaging: Gl~229~B (Nakajima et al. 1995), G196-3~B (Rebolo et al. 1998),
and GG Tau Bb (White et al. 1999). All of these companions are brown dwarfs 
confirmed by spectroscopy and proper motion.
The first extra-solar planet candidate (Terebey et al. 1998)
has not yet been confirmed by spectroscopy or proper motion.

We started a ground-based search for planets around nearby young stars: 
Their planets should be relatively luminous (because nearby and young) 
and well separated from the star. In addition, nearby stars usually have 
large proper motion, so that one can decide after only a few years, 
whether a companion candidate is co-moving. Then, its mass can be 
better constrained than for a free-floating object, because age 
and distance of the primary is usually well-known.
We report here the first {\em ground-based} direct imaging 
detection of an extra-solar planet candidate.

\section {The star 1RXSJ104230.3$-$334014 = TWA-7}

With follow-up observations of unidentified ROSAT X-ray sources,
many new low-mass pre-main sequence stars were discovered (Neuh\"auser 1997 
and references therein). One of the regions, where new, young stars were
found among ROSAT sources is the TW~Hya association (TWA),
a group of 14 young stars (Rucinski \& Krautter 1983, 
de la Reza et al. 1989, Gregorio-Hetem et al. 1992, Kastner et al. 1997, 
Jensen et al. 1998, Webb et al. 1999, Hoff 1999, Sterzik et al. 1999)
that share the same proper motion and radial velocity.
Many of these stars are multiple systems.
Four TWA members were observed with Hipparcos, and the weighted mean distance 
is 55 pc. Hence, TWA is the nearest association of young stars. 
Lowrance et al. (1999) and Webb et al. (1999) found a faint object $2^{\prime \prime}$ north
of the TWA member CoD$-33^{\circ}7795$. From its magnitudes and colors, they
concluded that the faint object may be a $\sim 20$~M$_{\rm jup}$ mass companion, 
but it is still unclear whether this visual pair forms a proper motion pair.

The apparent angular diameter of this association as measured from the largest
distance between two members is $30^{\circ}$, which corresponds to a projected
extent of 31 pc at the mean Hipparcos distance of 55 pc.
If the association has the same extent in radial direction, the distances of
the stars can range from 40 to 71 pc, i.e. slightly more than the range in 
distances of the four stars observed by Hipparcos. Thus, we can assume 
$55 \pm 16$ pc as distance of TWA members not observed by Hipparcos.

One of the new members of this association is 1RXSJ104230.3$-$334014
(Voges et al. 1999), also called TWA-7 (Webb et al. 1999). 
Using the fibre-fed spectrograph FEROS on the ESO 1.52m telescope, 
we obtained a spectrum of TWA-7 
on 1 June 1999 (3500\AA~to 9200\AA~with $\lambda/\Delta\lambda=48000$), 
see Fig. 1. 
TWA-7 is clearly a pre-main sequence star, as the equivalent width 
of the Li 6707\AA~line of this M1 star is $0.538 \pm 0.013$\AA . 
The barycentric radial velocity of TWA-7 is $11.80 \pm 0.29$~km/s,
consistent with the other TWA members.
Furthermore, we find W$_{\lambda}$(H$\alpha$)=$-5.23 \pm 0.10$\AA , 
i.e. TWA-7 is a weak-line T~Tauri star; its rotational velocity of 
v$\cdot \sin$~i =$6.1 \pm 0.5$ km/s is relativelly small.

\begin{figure}
\vbox{\psfig{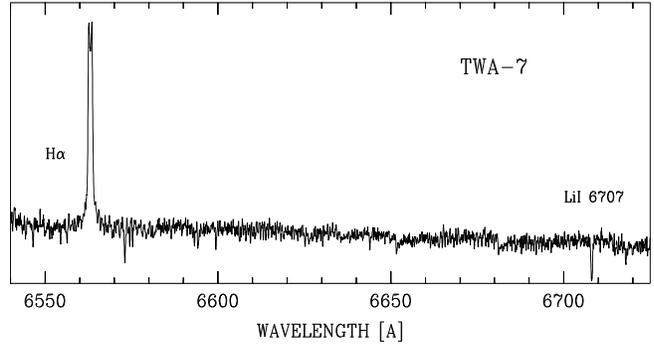}}
\caption{ Our high-resolution spectrum of TWA-7 shows strong lithium
absorption and H$\alpha$ emission, typical for young stars}
\end{figure}

TWA-7 is listed in the USNO-A2.0 (Monet et al. 1998) 
at $\alpha _{2000} = 10^{h} 42^{m} 30.28^{s}$ and 
$\delta _{2000} = -33^{\circ} 40\arcmin 16\farcs0$ for epoch 1982.235, 
with B~$\simeq 12.3$, and R~$\simeq 11.2$ mag, and also in the Hubble Space 
Telescope (HST) Guide Star Catalog (GSC1.2, Lasker et al. 1996) 
with B=$12.55 \pm 0.40$ mag.
In the STARNET catalog of proper motions and positions (R\"oser 1996),
the proper motion of TWA-7 on the Hipparcos system is given as
$\mu _{\alpha} = -136 \pm 5$ and $\mu _{\delta} = -19 \pm 5$ mas/yr
(S. Frink, priv. comm.). In the SPM catalog, Webb et al. (1999) found
$\mu _{\alpha} = -120 \pm 8$ and $\mu _{\delta} = -27 \pm 8$ mas/yr.
Because TWA-7 shares the radial velocity and proper motion
of the other TWA members, it is most certainly also a member.

From the spectral type M1 and the JHK photometry (Webb et al. 1999), we 
can conclude that the absorption is very small. Then, we obtain the bolometric 
luminosity at $55 \pm 16$ pc to be $\log$~L$_{\rm bol}$/L$_{\odot}=-0.41 \pm 0.13$.
A spectral type of M1 ($\pm$ one sub-type) corresponds
to T$_{\rm eff} = 3705 \pm 150$ K according
to Luhman (1999) for young M-dwarfs with surface gravities intermediate
between giants and dwarfs. 
The location of TWA-7 in the H-R diagram compared to evolutionary tracks 
and isochrones by D`Antona \& Mazzitelli (1994) and Baraffe et al. (1998) 
yields an age of 1 to 6 Myrs, i.e. co-eval with the other TWA stars,
and a mass of $0.55 \pm 0.15$~M$_{\odot}$.

\section {Infrared imaging}

TWA-7 was observed on 26 March 1998 with HST NIC2 (Near-Infrared Camera and 
Multi-Object Spectrometer, NICMOS) in coronographic mode with a F160W filter 
(as part of GTO 7226 by PI E. Becklin). 
Between two 224s exposures, 
HST was rotated by $29.9^{\circ}$ in order to facilitate the subtraction 
of instrumental signatures. We retrieved the individual pipeline calibrated 
images from the archive and subtracted them from each other to remove the 
diffraction pattern and the scattered light halo around TWA-7 (see Fig. 2).

A faint object south-east of TWA-7 clearly stands out on the subtracted data
(designated 1RXSJ104230.3$-$ 334014B = TWA-7B).
Centroiding TWA-7 behind the coronographic mask is problematic,
because the mask itself is not symmetric and is shifting its position by
up to $\pm 0.25$ pixel within one orbit. We fitted and extrapolated
the diffraction spikes to obtain their central crossing, which can be 
taken as the approximate location of the centroid of the occulted star,
and obtained a separation of $2.445 \pm 0.035^{\prime \prime}$
at a position angle of $142.24 \pm 1.34^{\circ}$.
The F160W magnitude of TWA-7B is $16.78 \pm 0.10$.

\begin{figure*}
\vbox{\psfig{figure=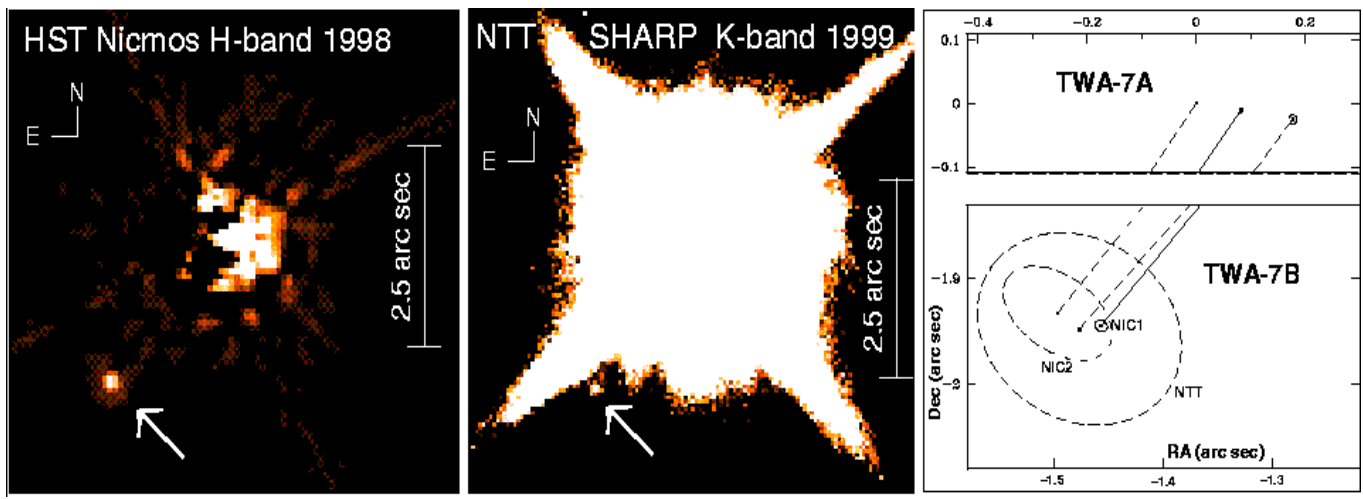,width=18cm,height=6.5cm,angle=270}}
\caption{ Images of TWA-7 with the faint object TWA-7B to the south-east: 
HST NICMOS F160W filter ($\sim$ H-band) where the bright star is partly occulted by 
the coronograph (left) and our NTT SHARP K-band detection (middle), we also detected
TWA-7B in H using SHARP. In the right panel, we show the 1$\sigma$ error ellipses of the 
positions of TWA-7B relative to TWA-7A at the three different epochs, taking into 
account the known proper motion of TWA-7 (here, we use the STARNET proper motion, 
but the result is basically the same when using the SPM proper motion),
first NIC2 (1998.2), then NIC1 (1998.8), then NTT (1999.5)
(upper right, from left to right, with small error circles in the TWA-7A positions 
due to error in proper motion).
For TWA-7B, the small innermost error ellipse is for NIC1 and the large outermost
one for NTT.
Due to the large astrometric errors in the NIC2 and NTT observations,
TWA-7B could be either a background object or a companion to TWA-7A}
\end{figure*}

On 19 June 1999, we observed the pair using SHARP (System for High 
Angular Resolution Pictures, Hofmann et al. 1992) at the ESO 3.5m New 
Technology Telescope (NTT) before the main targets of that program 
(63.N-0178) became visible at low airmass. 

The north-south alignment and the pixel scale of the camera were measured 
using images of the galactic center taken in the same night and precise
radio positions of those stars (Menten et al. 1997). We found the orientation 
to be off by $1.5 \pm 0.5^{\circ}$, namely tilted from N to W, 
and the pixel scale to be $0.04908 \pm 0.00064$ arc sec per pixel.

The SHARP speckle images consist of $1000 \times 0.5$s exposure in H 
and $500 \times 0.5$s in K, see Fig. 2. The short exposure 
seeing in the near-infrared during that night was better than $0.5^{\prime \prime}$.
The data were corrected for bad pixels followed by a sky image subtraction and
the application of a flat-field. For each band we then co-added the $256 \times 256$
pixel frames using the brightest pixel of TWA-7 as shift-and-add reference (Christou 1991).
We measure a separation between TWA-7 and 7B of $2.536 \pm 0.077^{\prime \prime}$ and,
after correcting for the misalignment, a position angle of 
$139.3 \pm 2.1^{\circ}$, consistent in both the H- and K-band image.

Using the standard star HR 4013 (Bouchet et al. 1991), 
observed just after TWA-7, we obtain 
H=7.11 and K=6.91 mag 
for TWA-7, within 0.02 mag of Webb et al. (1999).
For TWA-7B, we derive H=$16.42 \pm 0.11$ mag and K=$16.34 \pm 0.15$ mag,
i.e. more than nine magnitudes fainter than TWA-7.
These values are in agreement with the NICMOS data, especially considering 
the systematic offset between the HST Vega 
system\footnote{www.stsci.edu/instruments/nicmos/nicmos\_doc\_phot.html} 
and ground-based photometric systems for red objects of up to $0.2 \pm 0.1$ mag.

TWA-7 was again observed with HST NICMOS on 2 Nov 1998 (GTO 7226 by PI E. Becklin).
We found for TWA-7B the following magnitudes: $17.88 \pm 0.10$ in F090M (NIC1), 
$16.85 \pm 0.10$ in F165M (NIC2), and $16.9 \pm 0.2$ in F180M (NIC2). 
From the NIC1 image, the only one taken without coronograph,
we obtained $2.472 \pm 0.004^{\prime \prime}$ and $141.48 \pm 0.08^{\circ}$
for separation and position angle, respectively, between TWA-7 and 7B,
consistent with the other observations.

\section {Interpretation: Background or companion~?}

If TWA-7B orbits TWA-7, then they should form a common proper motion pair.
In Fig. 2 we show the 1$\sigma$ error ellipses of the positions of TWA-7B 
relative to TWA-7 as obtained at the three different epochs.
Taking into account the known proper motion of TWA-7, these
error ellipses overlap, consistent with TWA-7B being
a background object, which did not move relative to TWA-7.
However, due to the large errors in the NIC2 and NTT astrometry,
the data are also not inconsistent with TWA-7B being a companion.
Possible orbital motion of TWA-7B could
be in the direction opposite to the proper motion.
A final decision should be possible after a few more years.
The angular separation of $2.50 \pm 0.04^{\prime \prime}$ at a distance of $55 \pm 16$ pc 
corresponds to a projected separation of $138 \pm 37$ AU 
and an orbit period of $2185 \pm 1224$ yrs for a circular orbit.
 
The color of TWA-7B is H-K =$0.08 \pm 0.19$ mag, the absolute magnitudes
at $55 \pm 16$ pc are M$_{\rm H} = 12.72 \pm 0.76$ and 
M$_{\rm K} = 12.64 \pm 0.80$ mag (for negligible absorption). 
The errors in absolute magnitudes are mainly due to the 
error in distance, while apparent magnitudes are more precise. 
While the ground-based color does not strongly 
constrain the spectral type, the HST F090M-F165M-color 
($1.03 \pm 0.14$ mag) is consistent with a late K- or early M-type object, 
so that TWA-7B could be an unrelated background object. However,
the NICMOS quantum efficiency is very low in the F090M-filter (10 to 20$\%$)
leading to large and uncertain color terms, 
and this (intermediate-band, $\delta \lambda = 0.1885 \mu$m) filter
is more sensitive to spectral features than broad-band filters, 
so that the magnitude for this filter may be more uncertain.
Also, because a young planet should be variable in the infrared due to 
many impacts of planetesimals and comets, one should be careful with
comparing magnitudes obtained at different epoches.

The galactic model by Wainscoat et al. (1992) allows us to compute the density of 
sources in the direction of TWA-7. From their cumulative star counts 
for H$\le 16.5$ mag, we obtained a probability of $\sim 1~\%$
for finding one object within a $2.5^{\prime \prime}$ radius circle.
Even when considering that we observed four stars, where we found one such 
faint object within $2.5^{\prime \prime}$, the probability 
for TWA-7B being an unrelated background object is small.
Because the PSF of TWA-7B is consistent with a point-source rather than 
extended, it is unlikely that it is a background galaxy.  

According to 
Burrows et al. (1997), the ground-based H-K color is consistent 
with an object with effective temperature T$_{\rm eff} \simeq 1050$ K and surface 
gravity $\simeq 3000$~g/s$^{2}$. These values are consistent with an object 
with a mass of $\sim 3$M$_{\rm jup}$ and an age of $\sim 10^{6.5}$ yrs, 
which is also the age of TWA-7 and the other TWA members. With K=$16.34 \pm 0.15$ 
mag, $55 \pm 16$ pc distance, and B.C.$_{\rm K}$= 2 mag, we 
obtained a bolometric luminosity for TWA-7B of $\log$~L/L$_{\odot}=-4.0 \pm 0.3$.
An effective temperature of $\sim 1050$ K is similar to those of known
old T-dwarfs (Burgasser et al. 1999), so that the spectrum of TWA-7B should 
also show strong methane absorption features. 
If TWA-7B is a planet, would be just $\sim 3$ times more distant than 
the outermost jovian planet in the solar system and just $\sim 3$ times more
massive than the most massive planet in the solar system, so that such an object
may be less surprising than the very close-in 51 Peg-type planets.
 
Considering all arguments, it is possible that the HST and NTT images of TWA-7B 
presented here are the first direct images of an extra-solar planet, but
it seems more likely that this particular object
is an unrelated background object.
In any case, we have shown that {\em ground-based} direct imaging detection
of extra-solar planets at $\sim 100$~AU separation from a young star
is possible with current technology.

\acknowledgements{
We are grateful to E. Becklin for the HST GTO 7226 data, 
the referee M. McCaughrean for many helpful comments,
C. Leinert and J. Woitas for their help in determining 
the SHARP pixel scale and orientation, 
and S. Frink for providing the STARNET proper motion of TWA-7. 

{}


\begin{thebibliography}{}

\bibitem{} Baraffe I., Chabrier G., Allard F., Hauschildt P., 1998, A\&A 337, 403

\bibitem{} Bouchet P., Manfroid J., Schmider F.X., 1991, A\&AS 91, 409

\bibitem{} Brandner W., Alcal\'a J.M., Frink S., Kunkel M., 1997, The Messenger 89, 37

\bibitem{} Burgasser A.J., Kirkpatrick J.D., Brown M.E., et al., 1999, ApJ 522, L65

\bibitem{} Burrows A., Marley M., Hubbard W., et al. 1997, ApJ 491, 856

\bibitem{} Charbonneau D., Brown T.M., Latham D.W., Mayor M., 2000, ApJ, in press
(astro-ph/9911436)

\bibitem{} Christou J.C., 1991, Experimental Astr. 2, 27

\bibitem{} D'Antona F., Mazzitelli I., 1994, ApJS 90, 467

\bibitem{} Jensen E.L.N., Cohen D.H., Neuh\"auser R., 1998, AJ 116, 414

\bibitem{} Hoff W., 1999, PhD thesis, Universit\"at Jena

\bibitem{} Hofmann R., Blietz M., Duhoux Ph, Eckart A., Krabbe A., Rotaciuc V., 1992,
SHARP and FAST: NIR Speckle and Spectroscopy at the MPE. In: {\em Progress in Telescope 
and Instrumentation Technologies}, Ulrich M.-H. (Ed.), ESO Conference and Workshop 
Proceedings No. 42, 617

\bibitem{} Kastner J.H., Zuckerman B., Weintraub D.A., Forveille T., 1997, Science 277, 67

\bibitem{} Lasker B.M., Russel J.N., Jenkner H., 1996, The Guide Star Catalog Version 1.2 

\bibitem{} Lowrance P.J., McCarthy C., Becklin E.E., et al., 1999, ApJ 512, L69

\bibitem{} Luhman K., 1999, ApJ, in press (astro-ph/9905287)

\bibitem{} Malkov O., Piskunov A., Zinnecker H., 1998, A\&A 338, 452

\bibitem{} Marcy G.W., Butler R.P., 1998, ARA\&A 36, 57

\bibitem{} McCaughrean M.J., O'Dell C.R., 1996, AJ 111, 1977

\bibitem{} Menten K.M., Reid M.J., Eckart A., Genzel R., 1997, ApJ 475, L111

\bibitem{} Monet D., Bird A., Canzian B., et al., 1998, USNO-A2.0, 
A Catalog of Astrometric Standards, U.S. Naval Observatory Flagstaff

\bibitem{} Nakajima T., Oppenheimer B.R., Kulkarni S.R., et al., 1995, Nat 378, 463

\bibitem{} Neuh\"auser R., 1997, Science 276, 1363

\bibitem{} Rebolo R., Zapatero-Osorio M.R., Madruga S., et al., 1998, Science 282, 1309

\bibitem{} R\"oser S., 1996, In: IAU Symp. 172, 481

\bibitem{} Rucinski S.M., Krautter J., 1983, A\&A 121, 217

\bibitem{} Sterzik M.F., Alcal\'a J.M., Covino E., Petr M.G., 1999, A\&A 346, L41

\bibitem{} Terebey S., Van Buren D., Padgett D.L., Hancock T., Brundage M., 1998, ApJ 507, L71

\bibitem{} Voges W., Aschenbach B., Boller T. et al., 1999, A\&A 349, 389

\bibitem{} Wainscoat R.J., Cohen M., Volk K., Walker H.J., Schwartz D.E., 1992, ApJS 83, 111

\bibitem{} Webb R.A., Zuckerman B., Platais I., et al. 1999, ApJ 512, L63

\bibitem{} White R.J., Ghez A.M., Reid I.N., Schulz G., 1999, ApJ 520, 811

\end{thebibliography}
\end{document}